\documentstyle[12pt]{article}

\def\beq{\begin{equation}}

\def\eeq{\end{equation}}

\begin{document}
\pagestyle{empty}
$\ $
\vskip 1.5 truecm

\centerline{\bf INSTANTONS FOR VACUUM DECAY AT FINITE TEMPERATURE}
\centerline{\bf IN THE THIN WALL LIMIT}
\vskip .75 truecm
\centerline{ Jaume Garriga}
\vskip .4 truecm
\centerline{\em Tufts Institute of Cosmology,
              Department of Physics and Astronomy,}
\centerline{Tufts University, Medford, MA 02155}
\centerline{and}
\centerline{\em Department of Applied Mathematics and Theoretical Physics}
\centerline{Silver Street, Cambridge CB3 9EW, UK.}

\vskip 1. truecm

\begin{abstract}

In $N+1$ dimensions, false vacuum decay at zero temperature
is dominated by the $O(N+1)$ symmetric instanton, a sphere of radius
$R_0$, whereas at temperatures $T>>R_0^{-1}$, the decay is dominated by a
`cylindrical' (static) $O(N)$ symmetric instanton. We study the transition
between these two regimes in the thin wall approximation.
Taking an $O(N)$ symmetric ansatz for the
instantons, we show that for $N=2$ and $N=3$ new periodic solutions exist
in a finite temperature range in the neighborhood of $T\sim R_0^{-1}$.
However, these
solutions have higher action than the spherical or the cylindrical one. This
suggests that there is a sudden change (a first order transition) in the
derivative of the
nucleation rate at a certain temperature $T_*$,  when the static instanton
starts dominating. For $N=1$, on the other hand, the new solutions are
dominant and they smoothly interpolate between the zero
temperature instanton and the high temperature one, so the transition
is of second order. The determinantal prefactors corresponding to the
`cylindrical' instantons are discussed, and it is pointed out that the
entropic contributions from
massless excitations corresponding to deformations of the domain wall give
rise to an exponential enhancement of the nucleation rate for
$T>>R_0^{-1}$.

\end{abstract}

\clearpage
\pagestyle{plain}
\section{Introduction}
\label{Introduction}

Since the early work by Langer \cite{la67}, a lot of attention
has been devoted to the study of first order phase transitions using the
instanton methods \cite{co85}. These methods can be applied at
zero  temperature\cite{voal74,co85} as
well as at finite temperature \cite{af81},
and they have been widely used in connection with phase transitions in the
early Universe \cite{li83}. The basic predicament of this formalism is that the
false vacuum decay rate per unit volume is given by an expression of the form
\begin{equation}
{\Gamma\over V}=A e^{-S_E},
\label{basic}
\end{equation}
where $S_E$ is the Euclidean action of an instanton. The instanton is a
solution
of the Euclidean equations of motion with appropriate boundary conditions,
whose analytic continuation to Lorentzian time
represents the nucleation of a bubble of true vacuum in the false vacuum
phase. When more than one instanton is compatible with the
boundary conditions, one has to consider the one with the least Euclidean
action, which of course will dominate Eq. (\ref{basic}). The preexponential
factor $A$ arises from Gaussian functional integration over small fluctuations
around the instanton solution.

At zero temperature, the first order phase transition is dominated by
quantum tunneling through a potential barrier. This tunneling is
represented by a maximally symmetric instanton, which,
in $N+1$ spacetime dimensions, is a $O(N+1)$ symmetric configuration.
On the other hand, at sufficiently high temperature, the phase
transition is dominated by thermal hopping to the top of the potential barrier.
This process is represented by a `static' $O(N)$ symmetric instanton.

In the context of quantum mechanics (i.e., $0+1$ dimensions),
the decay of a metastable state at `intermediate' temperatures was studied
by Affleck \cite{af81} . Under certain assumptions for the shape of the
potential barrier, he found that the transition between quantum tunneling and
thermally assisted `hopping' occurred at a temperature $T_c$ of the order of
the curvature of the potential at the top of the barrier. At temperatures
$0<T<T_c$, the decay was dominated by `periodic' instantons (solutions of the
Euclidean equations of motion which are periodic in imaginary time), with
periodicity $\beta=T^{-1}$. Such periodic instantons smoothly interpolated
between the zero temperature instanton and the static `thermal hopping'
instanton.

However, as
noted by Chudnovsky \cite{ch91}, this situation is not generic. Depending on
the
shape of the potential barrier, it can happen that the transition from quantum
tunneling to thermal activation occurs at a temperature $T_*$ larger than
$T_c$, and in that case, there is a sudden discontinuity in the derivative of
the
nucleation rate with respect to the temperature at $T_*$. In \cite{ch91} this
was called a first order transition in the decay rate.

In the context of field theory, little work has been done toward the study of
periodic instantons other than the trivial static one. In this paper we shall
study such periodic instantons in the case when the so-called thin wall
approximation is valid. That is, when
thickness of the domain wall that separates the true from the false vacuum is
small compared with the radius of the nucleated bubble. Also, we shall restrict
ourselves to $O(N)$ symmetric solutions. As we shall see,
for $N=2$ and $N=3$ spatial dimensions, new periodic solutions exist but
only in a finite temperature range in the vicinity of $T\sim R_0^{-1}$, where
$R_0$ is the radius of the zero temperature instanton. We shall see that these
new instantons have higher Euclidean action than the static or the spherical
ones, and so they will never give the dominant contribution to the nucleation
rate (\ref{basic}). Barring the possibility that new non-$O(N)$-symmetric
instantons may alter the picture, this suggests that there is a sudden change
in the derivative of the nucleation rate with respect to the temperature
at $T_*$, the temperature at  which the action of the spherical instanton
becomes equal to the action of the cylindrical one. In the language of Ref.
\cite{ch91} this corresponds to a first order transition in the decay rate.
The case $N=1$ is somewhat different, since in that case the new
periodic solutions have lower action than the static one, and they mediate a
smooth transition between the zero temperature instanton and the high
temperature one.

Another point that we shall be concerned with is the evaluation of the
determinantal  prefactor $A$ for the case of the static instanton. In this
case,
the action is given by $S_E=-\beta E$, where $E$ is the energy of a critical
bubble (i.e. a bubble in unstable equilibrium between expansion and
contraction) and $\beta$ is the inverse of the temperature. As pointed out in
Ref. \cite{glal93}, apart from a factor of $T^{N+1}$ needed on dimensional
grounds, the prefactor $A$ is essentially the partition function of
small excitations around the critical bubble configuration. Upon
exponentiation, this has the effect of replacing $-\beta E$ by $-\beta F$ in
the
exponent of (\ref{basic}), where $F$ is the free energy of the bubble plus
fluctuations. At temperatures $T\sim R_0^{-1}$ the energy $E$ and the free
energy $F$ are approximately the same, and so the prefactor $A$ does not play
much of a role.  However, as we shall see, for $T>>R_0^{-1}$ and $N>1$, the
entropic contribution due to deformations of the domain wall can be very
large, giving an exponential enhancement to the decay rate.

The paper is organized as follows. In Section 2 we study the periodic
instantons, and the transition from the low to the high temperature regimes. In
Section 3 we evaluate the determinantal prefactors  for the static instanton in
the zero thickness limit, that is, the only fluctuations we consider are
deformations in the shape of the wall. This is supposed to be a good
approximation at temperatures $R_0^{-1}<<T<<m$, were $m$ is the mass of the
free particle excitations in the theory (supposed to be of the same order of
magnitude than the inverse thickness of the wall). Section 4 extends the result
for the prefactors to temperatures $m<<T$. Finally, some conclusions are
summarized in Section 5.

\section{The instantons}

In the thin wall approximation, the dynamics of a
bubble is well described by an action of the form
\begin{equation}
S=-\sigma\int d^N\xi \sqrt{\gamma}+\epsilon\int d Vdt.
\label{actioncov}
\end{equation}
The first term is  the Nambu action, proportional to the area of the worldsheet
of the wall separating true from false vacua, where $\sigma$ is the tension of
the wall, $\xi^a$ is a set of $N$ coordinates on the worldsheet and $\gamma$ is
the determinant of the worldsheet metric. The second term is the
volume enclosed by the wall times the difference in vacuum energy
density between both sides of the wall, which we denote by $\epsilon$,
integrated over time.

In spherical coordinates the metric in $N+1$
dimensional flat space-time reads
$$
ds^2=-dt^2+dr^2+r^2d^2\Omega^{(N-1)},
$$
where $d^2\Omega^{(N-1)}$ is the line element on the $(N-1)-$sphere. With a
spherical ansatz, the worldsheet of the domain wall separating the true
from the false vacuum is given by $r=r(t)$. In terms of $r$, the action
reads
\begin{equation}
S=-\sigma{\cal S}_{N-1}\left[\int r^{N-1}(1-\dot
r^2)^{1/2}dt-{\epsilon\over N\sigma}\int r^N dt\right], \label{action}
\end{equation}
where $\dot r=dr/dt$.
Here
$$
{\cal S}_{N-1}={2\pi^{N/2}\over\Gamma(N/2)}
$$
is the surface of the unit $(N-1)-$ sphere,  and we
have used ${\cal V}_{N-1}={\cal S}_{N-1}/N$, where ${\cal V}_{N-1}$ is the
volume inside the unit $N-1$-sphere.

Since the Lagrangian does not depend explicitly on time, the equation of
motion for the radius of the bubble, $r(t)$, reduces to the conservation
of energy,
\begin{equation}
E=p_r\dot r-L=\sigma{\cal S}_{N-1}\left[{r^{N-1}\over (1-\dot r^2)^{1/2}}
-{\epsilon\over N\sigma}r^N\right].\label{energy}
\end{equation}
This can be cast in the form
\begin{equation}
\dot r^2+V(r,E)=0,\label{particle}
\end{equation}
where
\begin{equation}
V=\left[{E\over \sigma {\cal S}_{N-1}}r^{1-N}+{\epsilon\over N\sigma}
r\right]^{-2}-1.\label{potential}
\end{equation}
Eq. (\ref{particle}) describes the motion of a non-relativistic particle
moving in the potential $V$.

To obtain the decay rates, one first has to find the relevant
instantons. These are solutions of the Euclidean equations of motion
with appropriate boundary conditions.
Taking $t\to -it_E$ in (\ref{particle}) we have
\begin{equation}
\left({dr\over dt_E}\right)^2-V=0. \label{particlee}
\end{equation}
At finite temperature $T$,
the instantons have to be periodic in $t_E$, with
periodicity $\beta\equiv T^{-1}$. Also, the solutions have to approach the
false vacuum at spatial infinity \cite{co85,af81,li83}.

The instanton for $E=0$ is well known. Integrating (\ref{particlee}) one
has
$$
r^2+t_E^2=R_0^2,
$$
where
$$
R_0={N\sigma\over\epsilon}.
$$
This is the zero temperature instanton, an $N$-sphere of radius $R_0$. It
has the $O(N+1)$ symmetry, as opposed to the solutions for $E\neq 0$, which
only have $O(N)$ symmetry. This instanton has Euclidean action
\begin{equation}
S_E^{0}={\sigma{\cal S}_N\over N+1}R_{0}^N.\label{se0}
\end{equation}
The same instanton also exists at finite temperature, $T<T_0\equiv
(2R_0)^{-2}$ (see Fig. 1a), except that the sphere is repeated periodically
along the $t_E$ axis. The action is still given by (\ref{se0}), and so it
is independent of temperature (apart from finite temperature corrections to
the parameters $\epsilon$ and $\sigma$, which are not relevant to our present
discussion).

For temperatures $T>T_0$ the spheres do not fit in the Euclidean time
interval, and the solution with $E=0$ no longer exists. It has been
suggested \cite{li83} that for $T>T_0$ the `overlapping' spheres would
somehow merge into a wiggly cylinder (see Fig. 1c), and ultimately
into a straight cylinder at even higher temperatures (Fig. 1b).
However, we would
like to stress that this is not necessarily so. As we shall see,
at least in the thin wall approximation and for $N$=2 and 3,
wiggly cylinder solutions exist only for a finite range of temperatures,
which happens to be below $T_0$ [see Eq. (\ref{range})].

Consider the case $N=2$. The corresponding potential $V$ is plotted in
Fig. 2. The
turning points are given by
$$
r_{\pm}={R_0\over 2}\left[1\pm\left(1-{4E\over \sigma R_0{\cal S}_{N-1}}
\right)^{1/2}\right].
$$
For $E>0$ we have Euclidean solutions which oscillate between the two
turning points. These are the wiggly cylinder solutions mentioned before.
The period of oscillation
$$
\tau(E)=2\int^{r_+}_{r_-}{dt_E\over dr}dr
$$
can be computed from (\ref{particlee}),
$$
\tau(E)=2[r_+{\cal E}(q)+{R_0^2\over r_+}{\cal K}(q)],
$$
where $q=(r_+^2-r_-^2)^{1/2}/r_+$ and ${\cal E}$ and ${\cal K}$ are the
complete Elliptic integrals \cite{grry}. A plot of $\tau(E)$ is given in
Fig. 3. By changing the energy, the periodicity $\tau(E)$ changes over a
finite range, and since $\tau=T^{-1}$, the wiggly cylinder solutions only
exist for a finite range of temperatures [see (\ref{range})].
The case $N=3$ is qualitatively similar. The equation for the turning
points, $V(r)=0$,
$$
{E\over \sigma {\cal S}_{N-1}}r^{1-N}+{r\over R_0}=1,
$$
has now three solutions, but one of them is negative, so again we have two
physical turning points.

In both cases, as the energy is increased from $E=0$, the two turning
points approach each other, and they eventually merge when
\begin{equation}
E=E_c\equiv {r_c^{N-1}\over N}\sigma  {\cal S}_{N-1}
\label{ec}
\end{equation}
at radius
$$
r=r_c\equiv {N-1\over N}R_0.
$$
These equations can be obtained from the conditions
\begin{equation}
V(r_c,E_c)=V'(r_c,E_c)=0,\label{conditions}
\end{equation}
where $V'=dV/dr$.
The period of the solutions as $E\to E_c$ is given by $\tau(E_c)=
2\pi/\omega_c$.
Here
$$
\omega_c^2={1\over 2}V''(r_c,E_c)
$$
is the frequency of the oscillations at the bottom of the well. Using
(\ref{potential}) we find
$$
T_c\equiv\tau^{-1}(E_c)={N\over(N-1)^{1/2}}(2\pi R_0)^{-1},
$$
hence the lower bound on the temperature range in which the wiggly
cylinder solutions exist
\begin{equation}
T_c<T<T_0. \label{range}
\end{equation}

It is clear from the conditions (\ref{conditions}) that the straight
cylinder (see Fig. 1)
$$
r=r_c
$$
is also a solution of (\ref{particlee}) corresponding to $E=E_c$. Since it
is static, this solution exists for all temperatures. The corresponding
Euclidean action
$$
S_E^c=E_c\beta,
$$
is inversely proportional to the temperature. Therefore this will be the
dominant instanton at sufficiently high temperature.

The physical interpretation of the three types of instantons can be inferred
from their analytic continuation to real time. The `straight cylinder'
represents thermal hopping to the top of the barrier, creating a critical
bubble in unstable equilibrium that can either recollapse under the tension
of the wall or expand under the negative pressure exerted by the false
vacuum. The `zero temperature' solution $E=0$ represents the spontaneous
quantum creation of a bubble from `nothing' (i.e. from a small quantum
fluctuation). This bubble is larger than the critical bubble and after
nucleation it expands to infinity. The wiggly cylinder instantons represent
a combination of both phenomena. A subcritical bubble of finite energy that
has been created by a thermal fluctuation tunnels through the barrier to a
supercritical bubble which can subsequently expand to infinity.

Let us now show that for $N=2$ and $3$ the process represented by the
wiggly cylinders is never dominant. From (\ref{energy}), Euclideanizing
and integrating over one period of oscillation ($S\to iS_E,t\to-it_E,
p_r\to ip_r^E$) we have ($S_E^w$ is the action for the wiggly cylinders)
$$
S_E^w=W+E\tau(E),
$$
where $W=\int p_rdr$. Using
$$
p_r^E=\sigma{\cal S}_{N-1}r^{N-1}{dr\over dt_E}\left[1+\left({dr\over
dt_E}\right)^2\right]^{-1/2}
$$
and (\ref{particlee}), one easily arrives at
$$
{dW\over dE}=-\tau(E),
$$
from which one obtains the Hamilton-Jacobi equation
\begin{equation}
{dS_E^w\over d\tau}=E\label{HJ}.
\end{equation}
As $E$ approaches $E_c$ from below, the amplitude of the wiggles tends to
zero and we have
\begin{equation}
\lim_{E\to E_c} S_E^w=S_E^c(E_c)\label{lim1}
\end{equation}
where $S^c_E(E_c)$ is the action for the straight cylinder at
$\beta=\tau(E_c)$. Also, when $E$ approaches zero from above the amplitude
of the wiggles becomes so large that in fact the cylinder fragments into
spheres
\begin{equation}
\lim_{E\to 0}S^w_E=S^0_E.\label{lim2}
\end{equation}
{}From (\ref{HJ}) one has
$$
0={dS_E^0\over d\tau}\leq{dS_E^w\over d\tau}\leq{dS_E^c\over d\tau}= E_c.
$$
It then follows that in the whole
range $0<E<E_c$ in which the wiggly cylinders exist, the action for the
wiggly cylinder $S_E^w$ is larger than both $S_E^c$ and $S_E^0$. This means
that the decay rate, which has the exponential dependence
$$
\Gamma\propto e^{-S_E}
$$
will be dominated either by the zero temperature instanton or by the
straight cylinder instanton, but never by the wiggly cylinder. The
situation is depicted in Fig. 4.

Define $T_*$ as the temperature at which the action for the cylinder is equal
to the action for the sphere, $S_E^c(\beta_*)=S_E^0$. For $N=2$
$\beta_*=(8/3)R_0$, and for $N=3$  $\beta_*=(27\pi/32)R_0$, so
$T_c<T_*<T_0$. For $T<T_*$ the decay is  dominated by the spherical
instantons, corresponding to zero temperature  quantum tunneling, whereas
for $T>T_*$ it is dominated by the cylindrical  one, corresponding to thermal
hopping. Since the slopes $$
{dS_E^c\over d\beta}\neq{dS_E^0\over d\beta}
$$
do not match at $T=T_*$, there will be a sudden change in the derivative of
the nucleation rate at this temperature. In the language of Ref.
\cite{ch91}, this corresponds to a first order transition in the
decay rate. This applies in the case when the thin wall approximation is valid.
To what extent the results apply for thick walls remains unclear
to us. Some comments  will be made in the concluding section.

The above considerations do not cover the case $N=1$. This case is special
because the force exerted by the false vacuum on the ``wall'' cannot be
compensated by the spatial curvature of the wall, which now reduces to a
pair of ``point-like'' kink and antikink. This case has previously been
considered in great detail in Ref. \cite{ivme87}, in the context of tunneling
and activated motion of a string accross a potential barrier.
For $E=0$ the solution of (\ref{particle}) is a circle of radius
$R_0=\sigma/\epsilon$, and so the zero temperature instanton is analogous
to the higher dimensional ones. For $E\neq 0$, however, the situation is
different because the potential $V$ has only one turning point. As a
result in the thin wall limit the periodic solutions, which are essentially
arcs of the circle of
radius $R_0$, contain mild singularities, vertices where the worldlines of
the kinks meet (see Fig. 5. Note that such vertices cannot be constructed
for $N>1$, since the worldsheet curvature would diverge there). The use of
vertices is just an artifact of the thin wall
approximation. They can be `rounded off'' by introducing a short range
attractive force between the kink and anti-kink, in the manner described in
\cite{ivme87}.  Because of this force, the potential $V$ is modified and a
new turning point appears at $r\sim\delta$, where $\delta$ is the range of
the attractive force (which in field theory will be of the order of the
thickness of the kink). As a result the instantons have a shape similar to
Fig. 1c.

For $N=1$ the zero temperature action is given by (\ref{se0}),
$S_E^0=\pi R_0\sigma$.
The action for the finite temperature periodic instantons
is $S_E^w=2\sigma l-\epsilon A$, where $l$ is the length of
one of the world-lines of kink or anti-kink and $A$ is the area enclosed
between them. We have
$$
S_E^w=\sigma R_0[\alpha+\sin \alpha],
$$
where $\alpha$ is related to the temperature $T=\beta^{-1}$ through (see
Fig. 5)
$$
\alpha=2\arcsin{\beta\over 2R_0}.
$$
For $\alpha=\pi$ this instanton reduces to the zero temperature one, the
circle of radius $R_0$.
As we increase temperature  above $(2R_0)^{-1}$, $\alpha$ decreases,
and in the limit
$\alpha\to 0$ we have
$$
S_E^w=2\sigma\beta+O(\alpha^3).
$$

In addition, if we take into account the short range attractive force
between the kink and anti-kink we will have the static instanton
for all temperatures  (the analogous of the straight cylinder representing
thermal
hopping). This consists of two parallel world-lines separated by a distance
$\sim\delta$, representing a pair in unstable equilibrium between the
short range attractive force and the negative pressure of the false vacuum
that pulls them apart \cite{prvi92}.
For $\delta<<R_0$ we can neglect the false vacuum term in
the action, and the action is just
proportional to the length of the world-lines
\begin{equation}
S_E^c\approx 2\sigma\beta.\label{wliness}
\end{equation}
In the high temperature limit this coincides with $S^w_E$.
The diagram of action versus temperature is sketched in Fig. 6.
The ``static'' instanton, whose action is depicted as a slanted solid line,
is always subdominant.
Since
$$
{dS_E^0\over d\beta}= {dS_E^v\over d\beta}=0,
$$
at $T=T_0=(2R_0)^{-1}$, it follows that the transition from the quantum
tunneling regime to the thermally assisted regime is now of second order
(in the terminology of \cite{ch91}).

Here we have concentrated in studying the instantons on the thin wall limit. It
is often claimed that the action for these instantons gives the exponential
dependence of the nucleation rates. However, as we shall see, contributions
from the determinantal prefactor can be exponentially large at temperatures
$T>>R_0$.

\section{Prefactors}

At zero temperature, the nucleation rates
per unit volume $\Gamma/V$ in the thin wall limit
have been given e.g. in \cite{ga93} (see also \cite{st76,af79,GNW80,afma82}).
We have
$$
{\Gamma\over L}={\epsilon\over 2\pi}e^{-S_E^0},\quad (N=1)
$$
$$
{\Gamma\over V}=\left({2\sigma R_0^2\over 3}\right)^{3/2} (\mu R_0)^{7/3}
R_0^{-3}e^{-S_E^0}, \quad (N=2)
$$
$$
{\Gamma\over V}=\left({\pi\sigma R_0^3\over 4}\right)^2{4R_0^{-4}\over
\pi^2}e^{\zeta'_R(-2)}e^{-S^0_E}, \quad (N=3)
$$
where $\zeta_R$ is Riemann's zeta function.
For $N=2$ the result contains an arbitrary renormalization scale
$\mu$ (see \cite{ga93}). These results should apply as long as
$T<min\{T_*,T_0\}$, when the decay is dominated by the spherical instanton.

For $T>T_*$ and $N=2,3$, as we have seen, the decay is dominated by the static
instanton. For an action of the form (\ref{actioncov}), the second variation of
the action due to small fluctuations of the worldsheet around a classical
solution is given by \cite{gavi92,ga93,jemal1}
$$\delta S^{(2)}={\sigma\over 2}\int d^N\xi\sqrt{\gamma}\phi\hat O\phi,$$
Physically, $\phi(\xi^a)$ has the meaning of a small normal displacement
of the worldsheet. Here $\xi^a$ is a set of coordinates on the instanton
worldsheet, $\gamma$ is the determinant of the worldsheet metric and
\begin{equation}
\hat O\equiv -\Delta_{\Sigma}+M^2,\label{op}
\end{equation}
where $\Delta_{\Sigma}$ is the Laplacian on the worldsheet and
$$M^2={\cal R}-\left({\epsilon\over\sigma}\right)^2,$$
whith ${\cal R}$ the intrinsic Ricci scalar of the worldsheet.

For the
static `cylinder' instantons,
$\Delta_{\Sigma}=\partial^2_{t_E}+\Delta^{(N-1)}$,
where $\Delta^{(N-1)}$ is the Laplacian on the $(N-1)-$sphere. The eigenvalues
of $\hat O$ in this case are given by
$$
\lambda_{L,n}=\left({2\pi n\over \beta}\right)^2+w_L^2,
$$
where
$$
w^2_L=(J^2-N+1)r_c^{-2},
$$
$J=L(L+N-2)$, $(L=0,1,...,\infty)$ and the degeneracy of each eigenvalue is
given by $g_L\cdot(2-\delta_{n0})$, where
$$
g_L={(2l+N-2)(N+L-3)!\over L!(N-2)!}.
$$
Since the critical bubble is a static configuration, it turns out that
$w^2_L$ are also
the eigenvalues of the second variation of the Hamiltonian
around this configuration. Note that the lowest eigenvalue, $L=0$, is negative
$w_0^2=-(N-1)r_c^{-2}$. This is because the static cylinder is in unstable
equilibrium between expansion and contraction.

The preexponential factor in the decay rate is given in terms of the
determinant of the operator $\hat O$ \cite{co85,af81,glal93,ga93}
\begin{equation}
\Gamma={|w_0|\over \pi}(\det\hat O)^{-1/2}e^{-S_E^c},
\label{reit}
\end{equation}
where
\begin{equation}
(\det\hat O)^{-1/2}=\prod_{L,n}(\lambda_{L,n})^{-(2-\delta_{n0})g_L/2}.
\label{det}
\end{equation}
For given $L$,
the product over $n$ can be easily computed by using the generalized
$\zeta-$function method \cite{ha77,ga93}. Actually, this product is just the
partition function for a harmonic oscillator with frequency $w_L$ at a
temperature $\beta^{-1}$,
\begin{equation}
Z_L=[2\sinh(\beta w_L/2)]^{-1},\label{zl}
\end{equation}
and we have
$$
(\det\hat O)^{-1/2}=\prod_L Z_L^{g_L}.
$$
For $L=1$, we have $w_1=0$ and so the previous expression is divergent. This is
due to the fact that some of the eigenmodes of $\hat O$ do not really
correspond to deformations of the worldsheet but to space translations of the
worldsheet as a whole. They are the so-called zero modes. As a result, $\hat O$
has $N$ vanishing eigenvalues, corresponding to the number of independent
translations of the cylinder (These eigenvalues are the $\lambda_{1,0}$, of
which there are $2 g_1=N$). The usual way of handling this problem is the
following. In the Gaussian functional integration over small fluctuations
around
the classical solution, which leads to the determinant in (\ref{reit}), one
replaces the integration over the amplitudes of the zero modes by an
integration over collective coordinates.

The transformation to collective
coordinates introduces a Jacobian
$$
J=\left({S_E^c\over 2\pi}\right)^{N/2},
$$
and the integration over collective coordinates introduces a space volume
factor $V$. The zero eigenvalues have now to be excluded from the determinant,
and noting that
$$
\lim_{w_1\to 0}{2\sinh(\beta w_1/2)\over w_1}=\beta,
$$
we have
\begin{equation}
(\det\hat O)^{-1/2}=\beta^{-N}VJ\prod_{L\neq 1} Z_L^{g_L}.
\label{detaqui}
\end{equation}
The decay rate at temperatures above $T_*$ is thus given by
\begin{equation}
{\Gamma\over V}={\beta^{-N}|w_0|\over 2\pi \sin(\beta w_0/2)}Je^{-\beta F},
\label{ratecorr}
\end{equation}
where
\begin{equation}
F\equiv E_c-\beta^{-1}\sum_{L>1}g_L\ln Z_L.\label{f}
\end{equation}
Here we have used $S_E^c=\beta E_c$, where $E_c$ is the energy of the critical
bubble, given by (\ref{ec}). Since $Z_L$ is the partition function for a
harmonic oscillator of frequency $w_L$ at temperature $\beta^{-1}$, the second
term in the r.h.s. of (\ref{f}) has a clear interpretation. Each physical mode
of oscillation will be excited by the thermal bath, and will make its
contribution to the free energy \cite{glal93}. Thus $F$ can be thought of as
the total free energy of the bubble plus fluctuations.

Let us analyze the contribution of the fluctuations to the free energy in more
detail. From (\ref{zl}) we have
\begin{equation}
-\ln Z_L={\beta w_L\over 2}+\ln(1-e^{-\beta w_L}).\label{fl}
\end{equation}
The first term in the right hand side contains the energy of zero point
oscillations of the membrane, and can be thought as quantum correction to the
mass of the membrane, of the same nature as the casimir energy which arises
in field theory in the presence of boundaries or non-trivial topology. Indeed,
from (\ref{op}), the fluctuations $\phi$ behave like a scalar field of
mass $M^2$ living on the worldsheet. Since the worldsheet has non-trivial
topology, this will result in a Casimir energy $\Delta E=\sum_{L>1}g_Lw_L/2$.
This expression is of course divergent and it needs to be regularized and
renormalized in the usual way.

For $N=2$ one obtains the renormalized
expression (see e.g. \cite{ka79})
\begin{equation}
\beta(\Delta  E)^{ren}={\beta\over 2}\left[-{1\over 6r_c}+{2\over
r_c}S(M^2r_c^2)+ M^2r_c\ln(\mu r_c)\right], \label{casimir}
\end{equation}
where
$$
S(a^2)\equiv \sum_1^{\infty}[(n^2+a^2)^{1/2}-n-a^2/2n].
$$
Notice the appearance of the arbitrary renormalization scale $\mu$ in the
previous expression. The coefficient of $\ln\mu$
can be easily deduced from the
general theory of $\zeta$-function regularizarion \cite{ha77,elal93}. It is
given by $-\zeta_{\hat O}(0)$, the generalized zeta function associated to the
operator $\hat O$, evaluated at the origin. This in turn can be expressed in
terms of geometric invariants of the manifold. In two dimensions one has
\begin{equation}
-\zeta_{\hat O}(0)={-1\over 4\pi}\int
d^2\xi^a\sqrt{\gamma}\left[-M^2+{1\over 6}{\cal R}\right],\label{zeta}
\end{equation}
where $M^2$ is the ``mass term'' in the Klein-Gordon operator $\hat O$ (in
general $M^2$ includes a linear coupling to the Ricci scalar,
of the form $\xi{\cal R}$).
In our case the Ricci scalar vanishes, and recalling that the
area of our instanton is given by $2\pi r_c\beta$ we recover the coefficient
of $\ln\mu$ in (\ref{casimir}).  From (\ref{zeta}), it is clear that a
redefinition of $\mu$ corresponds to a redefinition of the coefficient of the
Nambu term in the classical action, i.e., a redefinition of the tension
$\sigma$.
In our case $M=r_c^{-1}$,and  we have $(\Delta E)^{ren}\sim r_c^{-1}\ln(\mu
r_c)$.

For $N=3$, just on dimensional grounds $(\Delta E)^{ren}\sim r_c^{-1}$. In this
case there is no logarithmic dependence, since in odd dimensional spaces
without boundary $\xi_{\hat O}(0)$ always vanishes \cite{dona92}.
Comparing with the classical energy $E_c\sim \sigma r_c^{N-1}$, it is clear
that the Casimir energy is negligible as long as $\sigma r_c^N>>1$. In the
regime where the instanton approximation is valid, $S_E^c\sim\beta\sigma
r_c^{(N-1)}>>1$ , and since we are at temperatures $\beta{\buildrel
<\over\sim}r_c$, it follows that the Casimir term is negligible in front of the
classical one.

However, it is the second term in (\ref{fl}) that can be important, because of
its
strong temperature dependence. Note that this term, when summed over $L$
remains finite, and no further renormalization is needed. At temperatures
$T>>R_0^{-1}$, the mass $M$ is small compared with the temperature, and this
sum can be approximated by the free energy of a massless field living on the
surface of the membrane
$$
\sum_L g_L \ln(1-e^{-\beta w_L})\approx {\cal S}_{N-1}r_c^{N-1}\int d^{N-1}k
\ln(1-e^{-\beta k})+ O((T R_0)^{2-N}).
$$
In this limit
\begin{equation}
\beta F=\beta (E_c+\Delta E^{ren})-\alpha (r_cT)^{N-1}
\left[1+O\left((TR_0)^{-2}\right)\right].\label{conclude}
\end{equation}
Here $\alpha$ is a numerical coefficient
$$
\alpha={4\pi^{N-{1\over2}}\Gamma(N-1)\over
\Gamma(N/2)\Gamma\left({N-1\over 2}\right)}\zeta_R(N),
$$
with $\zeta_R$ the usual Riemann's zeta function.

To conclude, Eq.(\ref{conclude}) shows that at finite temperature $T>>R_0$, the
determinant of fluctuations around the critical bubble makes an exponentially
large contribution to the prefactor
\beq
e^{+\alpha(Tr_c)^{N-1}},  \label{pren1}
\eeq
which partially compensates the exponential suppresion due to the classical
action $e^{-\beta E_c}$. In particular, for $T\sim \sigma^{1/N}$ there is no
exponential suppression in the nucleation rate. This is to be expected, since
$\sigma^{1/N}$ is of the order of the Hagedorn temperature for the membrane.
(Of course, strictly speaking, our analysis of linear fluctuations of the
membrane will only be valid for $T<<\sigma^{1/N}$, since as we approach this
temperature thermal fluctuations of wavelength $\sim \beta$ become
non-linear).

For $N=1$ there is no exponential enhancement in the prefactor. As mentioned
before, the 1+1 dimensional case has been considered previously by Mel'nikov
and Ivlev \cite{ivme87}. They studied the prefactor accompanying the static
instanton which consists of a kink and antikink in unstable equilibrium between
their mutual short range attraction and the force exerted upon them by the
false vacuum, which tends to pull them apart. The temperature dependence of
the prefactor in this case is very easy to estimate in the thin wall limit.
There
are only two modes of fluctuation around the classical solution (as opposed to
the infinite number of modes for a string or a membrane), which
we shall take as $\phi_{\pm}\equiv \phi_1\pm\phi_2$. Here $\phi_1(t_E)$
and $\phi_2(t_E)$ are the
displacement of the kink  and the displacement of the
antikink from their equilibrium positions. It is clear that
$\phi_-$
will behave as an ``upside down'' harmonic oscillator, with imaginary frequency
$w_0$, ($w_0^2<0)$, which corresponds to the instability against increasing or
decreasing the distance between the kink and antikink. The variable $\phi_+$
on the other hand, behaves like a zero mode, corresponding to space
translations of the kink-antikink pair as a whole. The negative mode
$\phi_-$ will contribute
$$
Z_{w_0}=\left[2\sin{|w_0|\beta \over 2}\right]^{-1}
$$
to the prefactor. The zero mode $\phi_+$, which has to be treated in the
manner outlined before Eq. (\ref{detaqui}), will give rise to a factor
$\beta^{-1}\ell  J$, where $J=(S_E^c/2\pi)^{1/2}$, $\ell$ is the length of
space, and we have
$$
(\det\hat O)^{-1/2}=\beta^{-N}\ell JZ_{w_0}={\ell\over 2 \sin{|w_0|\beta\over
2}}\left({\sigma\over\pi\beta}\right)^{1/2}.
$$
Here $\sigma$ is just the mass of the kink, and we have used (\ref{wliness}).
Substituting in (\ref{reit}) we have
\begin{equation}
{\Gamma\over\ell}\approx {|w_0|\over 2\pi\sin{|w_0|\beta \over 2}}
\left({\sigma\over\pi\beta}\right)^{1/2}e^{-2\sigma\beta},\label{pre11}
\end{equation}
in agreement with \cite{ivme87}.

\section{Finite thickness}

So far we have treated the boundary that separates false from true vacua as
infinitely thin. However, in realistic field theories these boundaries always
have
some finite thickness $\delta$, and our estimates of the prefactor, Eqs.
(\ref{pren1}) and (\ref{pre11}), will have to be modified at temperatures
$T>>\delta^{-1}$. Here we shall briefly comment on such modifications
in the case when $\delta<<r_c$ (so that the thin wall aproximation is valid in
the
usual sense).

In general, we will have a scalar field (or order parameter) $\varphi$, whose
effective potential $V_{eff}(\varphi)$ has non-degenerate minima. For
definiteness we take the form
\begin{equation}
V_{eff}={\lambda\over 2}(\varphi^2-\eta^2)^2+{\epsilon\over 2\eta}\varphi.
\label{eff}
\end{equation}
It should be noted that $V_{eff}$ contains quantum corrections as well
as finite temperature corrections from all fields that interact with $\varphi$,
except from the field $\varphi$ itself \cite{glal93} and therefore the
parameters $\lambda$, $\eta$ and $\epsilon$ are temperature dependent
(this dependence is always implicit in what follows). From $V_{eff}$ one solves
the classical field equations to find the relevant instantons. Integration over
small fluctuations around the instantons yields the determinantal prefactors.
(For a detailed account of the formalism for the
computation of prefactors for field theoretic vacuum decay at finite
temperature, see Ref. \cite{glal93}.)

Among the modes of fluctuation around the static instanton, some correspond
to deformations of the membrane. These modes we have already considered in
the previous section. In addition, there will be
other eigenvalues of the fluctuation operator which correspond to
massive excitations of the scalar field $\varphi$ (both in the false and the
true
vacua), and to internal modes of oscillation in the internal structure of the
membrane. In simple models, such as (\ref{eff}), the energy scale
of internal excitations is of the same order of magnitude that
the mass of free particles of $\varphi$, which we denote by
$m=2\lambda^{1/2}\eta$. Therefore, even in the case when the thickness of
the wall is small in comparison with $ r_c$, the Nambu action
that we have used can be seen only
as an effective theory valid for $T<<m$, when the massive modes are
not excited.

For $N=1$ it was shown in \cite{ivme87} that the dependence of the prefactor
in (\ref{pre11}) as $T^{3/2}$ changes to $T^{-1/2}$ when $T>>m$. Also, for
$N>1$
we expect that the exponential growth of the prefactor with temperature,
Eq. (\ref{pren1}), will be modified for $T>>m$. Following \cite{glal93}, the
basic
difference with the zero thickness case is that the second term in the
right hand side of Eq. (\ref{conclude}), corresponding to the free energy of
massless excitations of the wall, will be replaced by the full finite
temperature correction to the free energy of a field theoretic bubble. If
$\delta<<r_c$ we have
\begin{equation}
\beta F\approx \beta E_c+\beta {\cal S}_{N-1}r_c^{N-1}(\Delta\sigma).
\label{freethick}
\end{equation}
Here $E_c$ is given by (\ref{ec}), with $r_c=(N-1)\sigma/\epsilon$, where
$\sigma=(4/3)\lambda^{1/2}\eta^3$, and $(\Delta \sigma)$ is the finite
temperature correction to the wall tension due to fluctuations of the field
$\varphi$ itself. We ignore the quantum correction $\Delta E^{ren}$
to the energy of the critical
bubble, which will be small compared to the classical term. Also, there are
classical corrections to $E_c$ of order $(\delta/r_c)^2$ with respect to the
leading term, due to the fact that the wall is curved \cite{af79}. We shall
also
ignore these.

The correction $(\Delta\sigma)$ can now be easily estimated in an adiabatic
approximation, since the temperature $T>>m$ is much larger than the inverse
thickness of the wall $\delta\sim m^{-1}$. To leading order in the thickness
over the radius, one can ignore the curvature of the wall. Neglecting the
second
term in (\ref{eff}), the scalar field has the well known solution representing
a
planar wall in the $z=0$ plane
$$
\varphi_w(z)=\eta\tanh\left({z\over\delta}\right),
$$
with $\delta=\lambda^{-1/2}\eta^{-1}$. Then
$$
\beta(\Delta\sigma)=\int dz\int {d^N k\over (2\pi)^N}
[\ln\left(1-\exp[-\beta(k^2+V_{eff}''(\varphi_w))^{1/2}]\right)-
$$
$$
\ln\left(1-\exp[-\beta(k^2+m^2)^{1/2}]\right)].
$$
The second term in the integrand subtracts the contribution of the
false vacuum.

For $N=3$, we can expand the integrand in powers of $(\beta m)^2$.
The leading contribution is then
$$
\beta(\Delta\sigma)\approx  {T\over 24}\int dz\
[V''_{eff}(\varphi_w)-m^2]=-{ m^2 T\delta\over 12}.$$
Substituting in (\ref{freethick}) and then in (\ref{ratecorr}) we have,
\begin{equation}
{\Gamma\over V}\propto T^4 \exp\left(-{E_c\over T}+{2\over 3}\pi Tm
r_c^2\right).\quad (N=3) \label{tgran3}
\end{equation}
Therefore, for $T>>m$, the finite temperature entropic contribution to the free
energy causes a correction to the nucleation rate which is still exponentially
growing with $T$ [this has to be compared with the behaviour for $T<<m^2$, Eq.
(\ref{pren1}) which depended even more strongly on $T$].

A similar calculation for $N=2$ yields
$$
\beta(\Delta\sigma)\approx
{m\over 2\pi}[\ln(\beta^2m^2)-1]-{1\over 8\pi}\int dz
V''_{eff}\ln(V''_{eff}/m^2)
 $$
where in the last term $V''_{eff}$ is evaluated at $\varphi_w(z)$. This last
term
is independent of the temperature, whereas the first only depends
logarithmically on it. As a result we have
\begin{equation}
{\Gamma\over V}\propto T^3 (T/m)^{2r_cm}e^{-\beta E_c}. \quad (N=2)
\label{tgran2}
\end{equation}
Therefore, for $T>>m$, the entropic correction is no longer
exponentially growing with $T$, but since $r_c>>m^{-1}$, the
preexponential power law enhancement can still be quite significant.

\section{Summary and discussion}

We have studied periodic instantons for vacuum decay in the thin wall
approximation, using a spherical ansatz for the spatial sections. For the case
of one spatial dimension ($N=1$), the periodic instantons have lower action
than the static one (see Fig. 6), and they mediate a
smooth transition between the high and the low temperature regimes.

For $N=2$ and $N=3$, on the other hand, we have seen that although new
periodic instantons exist in a certain temperature range [see Eq.
(\ref{range})], their action is larger than that of the
`spherical' low temperature instanton or the `static' high temperature one. As
a result, the new periodic solutions never dominate the vacuum decay. The
transition from low to high temperature regimes occurs  at the temperature
$T_*$ (see Fig. 4) at which the action for the spherical instanton coincides
with
the action for the static one. Since the slopes of the action versus
temperature
for both types of instanton do not match at $T=T_*$, this is called a first
order
transition in the nucleation rate at $T_*$ \cite{ch91}. As mentioned before, we
have treated the domain walls that separate true from false vacuum as
infinitely thin membranes, but we believe that these results will hold as long
as the thickness of the wall is much smaller than the radius of the nucleated
bubble. Of course, this can only be confirmed by studying a field theoretic
model such as the one given by Eq. (\ref{eff}). This is left for further
research.

In the case when the thickness of the walls is comparable to the bubble radius
we do not expect these results to hold, and it may well be that the periodic
instantons mediate a smooth transition between low and high temperature
regimes, even for $N>1$. As an extreme case, we can consider the soluble
model discussed in Ref. \cite{FSM84}, with effective potential
$$
V_{eff}={m^2\over 2}\varphi^2\left[1-\ln{\varphi^2\over c^2}\right],
$$
where $m$ and $c$ are constants. This model is not too realistic because the
mass of the $\varphi$ particles in the false vacuum ($\varphi=0$) is infinite.
Also, the potential is unbounded from below, so that actually there
is no true vacuum. Nevertheless this provides a framework in which one can
study periodic instantons. The model is extreme in the sense that the bubbles
have no core. At zero temperature, the instanton has the shape of a Gaussian
$$
\varphi=c e^{-{1\over 2}m^2\rho^2+2},
$$
where $\rho^2=t_E^2+\vec x^2$ (for definiteness we consider 4 spacetime
dimensions). At high temperatures $T>T_c\equiv m/(
2^{1/2}\pi)$, the decay is dominated by the static instanton
$$
\varphi=c e^{-{1\over 2}m^2r^2+{3\over 2}},
$$
where $r^2=\vec x^2$. At intermediate temperatures $0<T<T_c$ one can find
periodic instantons \cite{FSM84}
$$
\varphi=c f(t_E) e^{-{1\over 2}m^2r^2+{3\over 2}},
$$
where $f(t_E)$ satisfies the equation
\begin{equation}
\left({df\over dt_E}\right)^2=m^2 f^2 (4-\ln f^2)+{2m^3 E\over \pi^{3/2}c^2}.
\label{jt}
\end{equation}
Here $E$ is the energy of the nucleated bubble. The period $\beta$ of the
solutions of (\ref{jt}) is plotted in Fig. 7 as a function of $E$ . Note that
the
period decreases with increasing energy, in contrast with Fig. 3, which
corresponds to the thin wall case. Also, as $E\to 0$ we have $\beta\to \infty$
and the action approaches that of the zero temperature instanton. For $E\to
E_c\equiv e^3\pi^{3/2} c^2/2m$, it approaches that of the
static one \cite{FSM84}.
Then, from the Hamilton-Jacobi equation (\ref{HJ}) and the fact that
$\beta(E)$ is monotonously decreasing, it follows that the action for the
periodic instantons is lower than that of the static or the zero temperature
ones in the whole range $0<T<T_c$. In that case, such periodic
instantons mediate a smooth transition between the low and high temperature
regimes.

Finally, we have studied the determinantal prefactor
in the nucleation rate for the case of the static instanton, in the thin wall
case.
We have seen that
for $N>1$ and $T>>R_0$ there is an exponential enhancement of the nucleation
rate [see Eq. (\ref{pren1})] which can be understood as the contribution to
the free energy of the fluctuations around the critical bubble. Deformations of
the membrane behave essentially as a scalar field of small mass (of order
$R_0^{-1}$) living on the worldsheet of the wall. The enhancement (\ref{pren1})
is basically the partition function for a gas of such excitations living on the
worldsheet.

At temperatures $T>>m$ (where $m$ is the mass of
free particles in the false vacuum, assumed to be of the same order of
magnitude than the inverse thickness of the wall), the exponential law
(\ref{pren1}) no longer applies, since one has to consider excitations in the
internal structure of the membrane. In this regime one obtains (\ref{tgran3})
and (\ref{tgran2}), which still contain a sizable
enhancement above the naive estimate $(\Gamma/V)\propto
T^{N+1}\exp(-\beta E_c)$.

The case $N=1$ had been considered in
\cite{ivme87}. In this case there is no exponential enhancement because the
membrane reduces to a pair of kink and antikink, so there is no `gas' of
massless excitations contributing to the free energy. For $R_0<<T<<m$ the
prefactor behaves as $T^{3/2}$ [see (\ref{pre11})], whereas for $T>>m$ it
behaves as $T^{-1/2}$ \cite{ivme87}.

\section*{Acknowledgements}

I would like to thank Alex Vilenkin for drawing my attention to this problem,
and to Tanmay Vachaspati for interesting conversations.
This work was partially supported by the National Research Foundation
under contract PHY-9248784. I am indebted to Julie Traugut for the layout
of Figs. 1 and 5.

\section*{Figure captions}
\begin{itemize}

\item{\bf Fig. 1} Different types of instantons are found in different
temperature regimes. Fig. 1a represents the spherical instanton, which exists
at zero temperature but also at any temperature lower than $(2R_0)^{-1}$. Fig.
1b represents the  `cylindrical' instanton, which, being static, exists for all
temperatures. Finally, Fig. 1c represents the new instantons, periodic `wiggly'
cylinders which exist only for a finite temperature range in the vicinity of
$T\sim R_0$. The periodicity is $\beta=T^{-1}$.

\item{\bf Fig. 2} Represented is the potential of Eq. (\ref{potential})
for $N=2$, as a function
of $r$. The solid line corresponds to an energy lower than $E_c$. Since in this
case the equation $V(r)=0$ has two solutions, the instanton has two turning
points. As the energy is increased to $E=E_c$ the shape of the potential
changes and the two turning points merge into one at $r_c$. The dashed line
represents the shape of the potential for $E=E_c$.

\item{\bf Fig. 3} The period $\tau$ of the `wiggly cylinder' solutions as a
function of energy. By changing the energy, the periodicity changes only over a
finite range, and so these solutions exist only for a finite range of
temperatures.

\item{\bf Fig. 4} The sketch represents the action for the different types of
instantons as a function of the inverse temperature $\beta$, for $N>1$. The
horizontal line represents the action for the spherical instantons, which is
independent of temperature in the range where it exists. The slanted straight
line is the action for the static instanton, which is proportional to $\beta$,
the
length of the cylinder. The dashed line represents the action for the wiggly
cylinders, which exist only in a finite temperature range.

\item{\bf Fig. 5} Periodic instantons for the case $N=1$, in the thin wall
limit.

\item{\bf Fig. 6} The action for the different types of
instanton in the case $N=1$. Same conventions as in Fig. 4. In this case, the
action of the periodic instantons (dashed line) is always smaller than the one
for the static instanton (solid slanted line). The periodic instantons smoothly
interpolate between the spherical instanton action (horizontal line) and the
high temperature static instanton action. The slope of the dashed line at
$\beta=2R_0$ vanishes.

\item{\bf Fig. 7} The period $\beta$ of the solutions of (\ref{jt}) as a
function of
energy $E$.

\end{itemize}

\end{document}